\documentclass[12pt]{article}
\usepackage{amssymb}
\usepackage{amsmath}
\usepackage{graphicx}
\usepackage[matrix,arrow]{xy}
\parskip=\medskipamount
\newtheorem{theorem}{Theorem}

\newtheorem{proposition}[theorem]{Proposition}

   \def\Ga{\Gamma}
   
\def\ep{\epsilon}

\def\om{\omega}   
   
\def\IC{\relax{\it  l\kern-.50 em C}}
\def\IE{\relax{\it  l\kern-.12 em E}}
\def\IK{\relax{\it  l\kern-.18 em K}}
\def\IL{\relax{\it  I\kern-.18 em L}}
\def\IN{\relax{\it  I\kern-.18 em N}}
\def\IR{\relax{\it  I\kern-.18 em R}}

\def\<#1>{\langle#1\rangle}
\def\d<#1>{\langle\langle#1\rangle\rangle}
\font\tenfrak=eufm10  \font\sevenfrak=eufm7  \font\fivefrak=eufm5
\newfam\frakfam
\textfont\frakfam=\tenfrak
\scriptfont\frakfam=\sevenfrak
\scriptscriptfont\frakfam=\fivefrak

\def\Diff{\mathop{\it  Diff}\nolimits}

\def\frac#1#2{{#1\over #2}}

\def\pd#1#2{\frac{\partial #1}{\partial#2}}

\def\forms{{\textstyle \bigwedge}}
\def\Ga{\Gamma}
\def\om{\omega}  

\def\Diff{\mathop{\rm Diff}\nolimits}
\def\vfield{{\mathfrak{X}}}

\begin{document}

\title{Canonoid transformations and master symmetries}

\author{
Jos\'e F. Cari\~nena$^{a)}$, Fernando Falceto$^{b)}$ and Manuel F.
Ra\~nada$^{c)}$
\\[4pt]
  {\sl Departamento de F\'{\i}sica Te\'orica, Facultad de Ciencias} \\
  {\sl Universidad de Zaragoza, 50009 Zaragoza, Spain}
}

\maketitle

\begin{abstract}
Different types of transformations of a dynamical system, that are compatible with the Hamiltonian structure, are discussed making use of a geometric formalism.  
Firstly, the case of canonoid transformations is studied with great detail and then the properties of  master symmetries are also analyzed.  The relations between
 the existence of constants of motion and  the properties of canonoid symmetries is discussed making use of a family of boundary and coboundary operators.
 \end{abstract}
\begin{quote}

{\sl Running title:}{\enskip}
Canonoid transformations and master symmetries



\end{quote}
{\vfill}

\footnoterule
{\noindent\small
$^{a)}${\it E-mail address:} {jfc@unizar.es} \\
$^{b)}${\it E-mail address:} {falceto@unizar.es}\\
$^{c)}${\it E-mail address:} {mfran@unizar.es}  }
\newpage

\section{Introduction}

In the search for the general solution of a dynamical equation one can use  an appropriate transformation 
of the given equation into a simpler one,
but one can also make use of reduction procedures leaving to  related simpler systems. Such  reduction processes are based on the 
determination  of constants of the motion, on one side, or infinitesimal symmetries of the dynamics, on the other.

 In differential geometric terms, 
 the dynamics is described by means of a vector field and therefore the theory of transformations of such vector
  fields is a very important geometric ingredient. In the above mentioned reduction processes the constants of the motion give rise 
  to invariant foliations. On the other side, infinitesimal symmetries of the dynamics allow us to introduce adapted coordinates. Then,  
  the system of differential equations splits into a simpler one involving one less coordinate
  and another single equation to be solved once the other subsystem has been solved. 
  
  The existence of additional structures compatible with the
   dynamics provides us with additional tools. In particular, a compatible symplectic structure gives us an identification of 
   vector fields with 1-forms, and therefore there is a distinguished class of vector fields, those associated with exact (or at least closed) 
   1-forms. Consequently, functions play the additional  role of being generators of  Hamiltonian vector fields. Noether theorem in Hamiltonian dynamics identifies
    constants of motion with generators of infinitesimal strictly canonical symmetries of the Hamiltonian, therefore the knowledge of a constant of the motion reduces the 
    problem to another one involving two less degrees of freedom. Of course functionally independent constants of the motion
     cannot be used simultaneously in this way unless they are in involution. 

 The main objective of this article is to develop a deeper analysis of the theory of transformations on symplectic manifolds. 
The paper is organized as follows. In Section 2, we introduce the notation and give a short review of the theory of  canonical transformations, non-strictly canonical 
transformations, and canonoid transformations using the symplectic formalism as an approach.  
Section 3 is devoted to the study of the one-parameter groups of master symmetries and canonoid symmetries and in Section 4 the relation with the existence of constants of motion is studied.  Finally, in Section 5 we make some final comments.  Appendix A summarizes some properties of two homological 
differential operates and Appendix B shows the possibility of choosing an appropriate 1-form used in Section 3.

\section{Transformations in symplectic manifolds}   
We first recall that a symplectic manifold is a pair $(M,\omega)$,  where  $M$ is a differentiable manifold 
endowed with a  symplectic form $\omega$, which is  a nondegenerate closed 2-form in $M$, 
 $d\omega = 0$, i.e. $\omega\in Z^2(M)$ (see e.g. \cite{AbrMarsden}-\cite{MoFelV90}). There is then a one-to-one $C^\infty(M)$-linear correspondence between the $C^\infty(M)$-module   of vector fields and that of 1-forms: If 
 $X$ is in the $C^\infty(M)$-module $\mathfrak{X}(M)  $ of vector fields in $M$,  the corresponding 1-form is denoted $\beta_X=i(X)\omega$, and if $\beta\in \forms^1(M)$, its associated 
 vector field $X_\beta$ is the one such that $i(X_\beta)\omega=\beta$. Those vector fields associated to closed 1-forms are  called 
locally-Hamiltonian vector fields and in particular vector fields associated with exact 1-forms $df$ are said to be Hamiltonian vector
 fields and are denoted $X_f$ instead of $X_{df}$. The set of 
locally-Hamiltonian and Hamiltonian 
 vector fields  are $\mathbb{R}$-linear spaces 
 to be denoted, respectively,  
 $\mathfrak{X}_{\rm LH}(M,\omega)$ and $\mathfrak{X}_{\rm H}(M,\omega)$.
  
A diffeomorphism $\Phi$ of a manifold $M$  push-forward tensorial fields in $M$. We use the notation  $\Phi_*$ instead of  $(\Phi^{-1})^*$ 
for covariant tensors. So, $\Phi_*f=(\Phi^{-1})^*f=f\circ\Phi^{-1}$, $\forall f\in C^\infty(M)$, while $\Phi_*(X)$ is obtained from $\Phi_*(X)(\Phi_*f)=\Phi_*(Xf)$,  $\forall f\in C^\infty(M)$.

Those diffeomorphisms  leaving invariant a particular tensor field  are called symmetries of such a tensor field. 
Next we consider three fundamental examples: 

\begin{enumerate}

\item  If $f$ is a function defined in $M$, $f\in C^\infty(M)$, then a symmetry of $f$ is a diffeomorphism $\Phi$ of $M$ such that $\Phi_*(f)=f$.

\item  If $X$ is a vector field on $M$, $X\in\mathfrak{X}(M)$,  then a symmetry of $X$ is a diffeomorphism $\Phi$ of $M$ such that $\Phi_*(X)=X$. 

\item  If $\alpha$ is a $k$-form in $M$, $\alpha\in\forms^k(M)$, then a symmetry of $\alpha$ is a diffeomorphism $\Phi$ of $M$ such that $\Phi_*(\alpha)=\alpha$.

\end{enumerate}

In this section we study, by making use of a geometrical approach, three different classes of transformations related with the properties of the Hamiltonian formalism: strictly canonical transformations, non-strictly canonical transformations, and canonoid transformations.

In Classical Hamiltonian Mechanics, those transformations of the phase space that preserve the Hamiltonian  form of the Hamilton equations, whatever the Hamiltonian function is, are called canonical.  These transformations are characterized by the existence of a real number $\lambda$, called valence, such 
that the Poisson bracket  of two transformed functions is $\lambda $ times the transformed of the Poisson bracket of   the original functions \cite{SaletanCromer,{SaletanJose}}.   The set of canonical transformations is endowed with 
 a group structure and the set of strictly canonical transformations, those corresponding to $\lambda=1$, is a normal subgroup.  In differential geometric terms the phase space is a symplectic manifold $(M,\omega)$  and strictly canonical transformations  are represented by diffeomorphisms $\Phi\in \Diff(M)$ that preserve the symplectic form, that is,  $\Phi_*(\omega)=\omega$.

\subsection{Strictly canonical transformations }   
\indent 

 In a symplectic manifold $(M,\omega)$ the symmetries of $\omega$, to be called   symplectomorphisms, 
 are diffeomorphisms of $M$ such that  $\Phi_*(\omega) = \omega$, what is equivalent to $ \Phi^*(\omega) = \omega$. 
   The remarkable point is that if $H$ is a Hamiltonian function and $\Ga_H$ is the associated 
    vector field representing the dynamics, i.e.  satisfying the equation
\begin{equation}
  i(\Ga_H)\omega = dH  \,, \label{Hdyneq}
\end{equation}
then the following equation is also true 
\begin{equation}
  i(\Phi_*(\Ga_H))\Phi_*(\omega) = d(\Phi_*(H))   \,.       \label{tlaw}
\end{equation}
Therefore symplectomorphisms that are symmetries of $H$ are also symmetries of $\Ga_H$. 
However, the above equation permits the existence of symmetries of  $\Ga_H$ that are not symplectomorphisms.
 Of course in this last case  the new 2-form $\Phi_*(\omega) \ne\omega $ is admissible for $\Ga_H$ and, as pointed out in \cite{CaR88},
  the vector field $\Ga_H$ turns 
out to be a bi-Hamiltonian system \cite{Mag78} and therefore non-Noether constant of motion can be found \cite{CI83}.

Given two functions  $f$ and $g$ in a symplectic manifold $(M,\omega)$,  then the symplectic product of the corresponding Hamiltonian vector 
fields represents the so called Poisson bracket of these two functions 
$$
 \{f,g\} = \om(X_f,X_g)=-X_fg=X_gf \,.
$$

 Note that for symplectomorphisms $\Phi$ of $M$, $\Phi_*X_f=X_{\Phi_*f}$, and this property leads  to 
 $$
 \Phi^*\{f,g\} = \{\Phi^*f,\Phi^*g\},
 $$
because 
$$
\{\Phi^*f,\Phi^*g\}=X_{\Phi^{-1}_*f}\Phi^*g=\Phi^{-1}_*(X_f)(\Phi^*g)=\Phi^*(X_fg)= \Phi^*\{f,g\},
$$
and therefore the symplectomorphisms preserve the  Poisson brackets of any pair of functions. Consequently, they correspond to strictly canonical transformations.

At the infinitesimal level, one-parameter subgroups of symmetry transformations of tensor fields are characterized by the vanishing of the Lie derivative of the tensor field with respect to the vector field $X$ generating the one-parameter subgroup, that is, (i)   $\mathcal{L}_Xf:=X(f)=0$  for  functions, (ii)  $\mathcal{L}_X\Ga:=[X,\Gamma]=0$ for  vector fields $\Gamma$, and in general (iii) $\mathcal{L}_X\alpha :=(d\circ i(\Ga)+i(\Ga)\circ d)\alpha= 0$ for a $k$-form $\alpha$.  
In particular, locally-Hamiltonian vector fields in a symplectic manifold are infinitesimal symplectomorphisms.
 
 Of course, when there exist tensorial relationships among tensorial objects their infinitesimal symmetries are also related.  Next we consider two particular situations.

 (1) Let us first consider the Lie derivative  with respect to a vector field $X$ of the dynamical equation (\ref{Hdyneq}) and use the property $\mathcal{L}_Xi(Y)\alpha-i(Y)\mathcal{L}_X\alpha=i([X,Y])\alpha$, $\forall X,Y\in\mathfrak{X}(M)$ and $\alpha\in\forms(M)$, and we find  
\begin{equation}
  \mathcal{L}_X\Bigl(i(\Ga_H)\omega-dH\Bigr) =  i(\Ga_H) \mathcal{L}_X\omega+i([X,\Ga_H])\omega-d(XH) = 0 \,. \label{Liederdyn}
\end{equation}
Then, if $X$ is an infinitesimal symmetry of both $\omega$ and $H$ (that is,  $\mathcal{L}_X\omega=0$ and $XH=0$)  we obtain 
$$
  i([X,\Ga_H])\omega = 0\,, 
$$
and, as $\omega$ is nondegenerate, this means that $X$ is a symmetry of the dynamical vector field $\Ga_H$.

(2)  Let us now suppose that $X$ is such that   $\mathcal{L}_X\omega=0$  and $\mathcal{L}_X\Ga_H=[X,\Ga_H]=0$. 

Then (\ref{Liederdyn}) shows that  $\mathcal{L}_XdH=0$, and therefore, if $M$ is connected  $\mathcal{L}_XH={\rm const.}$ 
In particular,  if a Hamiltonian vector field $X_f$  is a symmetry of $\Ga_H$  (but not of $H$) 
then $f$ is not necessary a constant of the motion since the vanishing of the Lie bracket  $[X_f,\Ga_H]=0$ only means the vanishing 
of the differential of the Poisson bracket 
$$ 
  d \bigl(\{f,H\}\bigr) = 0,
$$
and from here we can conclude when $M$ is connected  that $\mathcal{L}_{\Ga_H}f=\{f, H\}$ is a (not necessarily zero) constant \cite{LoMR99}.

Finally, let us mention that, given a Hamiltonian system $(M,\omega,H)$, one usually look for 
vector fields whose flows are symplectomorphisms that are also symmetries of $H$ and, therefore, symmetries of $\Ga_H$.
Then for each $g\in C^\infty(M)$, the relation 
 $$ 
  \mathcal{L}_{\Ga_H}g = \{g,H\} = -\,\mathcal{L}_{X_g}(H)
 $$
 shows that $X_g$ is a symmetry of $H$ if and only if $g$ is a constant of motion. This is a very important property,
 sometimes called Noether's theorem in Hamiltonian formalism, because it suggests us  a method for finding constants 
 of the motion which are very useful in the process of reduction of the dynamical equation.
 
 The usefulness of non-strictly canonical infinitesimal symmetries (see e.g a generalisation of the virial theorem that can be found in
the recent paper  \cite{CFR12}) and the more general case of canonoid transformations 
 has been less analyzed and is worthy of a deeper analysis. Several applications of canonoid transformations can be seen at \cite{LS07,TALS} (see also \cite{DTH} for the Nambu formulation).

\subsection{Non-strictly canonical transformations}
 \indent 
 
 As indicated above canonical transformations are those preserving the form of Hamilton equations whatever the Hamiltonian is, or in an equivalent way, preserving the Poisson bracket of any two functions up to a nonzero multiplicative constant: the valence.
A transformation with valence different from one is called
non-strictly canonical \cite{SaletanCromer}, while those with 
valence equal to one are said to be strictly canonical. 
In differential geometric terms these canonical transformations 
in a symplectic manifold $(M,\omega)$ are represented by 
diffeomorphisms $\Phi$ such that 
$$
  \Phi^*(\omega)=r\,\omega, \quad r\in\mathbb{R}\,,        \label{nsct}
$$
and strictly canonical ones are those with $r=1$.

Let the vector field $X$ be the generator of a one-parameter group of  canonical transformations $\Phi_\epsilon^*(\omega)=r(\epsilon)\,\omega$.  Then, there exists a real number $a\ne 0$ such that 
\begin{equation}
 \mathcal{L}_{X}\omega=a\,\omega ,\label{infnsct}
 \end{equation}
with $r$ and $a$ related by $r(\epsilon)=e^{a\epsilon}$. In an equivalent way, in terms of $\beta_X=i(X)\omega$, the canonicity condition (\ref{infnsct}) reads $d\beta_X=a\,\omega$.
In particular, the flow of $X$ is made up of strictly canonical transformations (symplectomorphisms) when $a=0$, i.e. when $\beta_X$ is closed.
 
A diffeomorphism $\Phi$ on $(M,\omega)$ such that 
$$ 
  \Phi^*(\omega)=r\,\omega \,,{\qquad} \Phi^*(H)=r\,H \,,\qquad {\rm with}\quad r\in\mathbb{R}, 
$$
preserves the Hamiltonian vector field $\Ga_H$, because of (\ref{tlaw}). 
At the infinitesimal level, if $X$ is such that $ \mathcal{L}_X\omega=a\,\omega$ and $X(H)=a\, H$, then 
 $$ 
i([X,\Ga_H])\omega =  \mathcal{L}_X[i({\Ga_H}) \omega] -i({\Ga_H})\mathcal{L}_X\omega = \mathcal{L}_X(dH)-a\,i(\Ga_H)\omega=d(a\,H)-a(dH)=0 ,
 $$
and, using that $\omega$ is nondegenerate, we arrive at
$$
  [X,\Ga_H] = 0 
$$
so that $X$ is a symmetry of the dynamical vector field.
 
 On the other side, if $X$ is such that  $ \mathcal{L}_X\omega=a\,\omega$, then
we have 
\begin{equation}
 i([X,\Ga_H])\omega  = ( \mathcal{L}_Xi(\Gamma_H)-i(\Gamma_H) \mathcal{L}_X)\omega=\mathcal{L}_X (dH) -a\,i(\Ga_H)\omega=d(\mathcal{L}_XH-aH). \label{condcansym}
\end{equation}
Therefore, when  $ \mathcal{L}_X\omega=a\,\omega$ and $M$ is connected,   $ [X,\Ga_H]=0$ if and only if $\mathcal{L}_XH-aH$ is  a numerical constant.

Let us choose a
vector field $X_1$  such that 
$$\mathcal{L}_{X_1}\omega=-\omega.
$$
This is only possible when $\omega$ is exact because $\omega=d(-i(X_1) \omega)$. 
For instance, when  $(M,\omega=-d\theta )$ is  an exact symplectic manifold  the vector field  $X_1$ can be chosen
 to be defined by (see \cite{NachVer86})
$$i(X_1)\omega=\theta,\qquad {\rm i.e.}\ X_1=X_\theta,$$ 
because then
$$\mathcal{L}_{X_1}\theta=i(X_1)d\theta+d(i(X_1)\theta)=-i(X_1)\omega +d(i(X_1)\theta)=-\theta+d(i(X_1)\theta),
$$
and therefore
$$\mathcal{L}_{X_1}\omega=-\mathcal{L}_{X_1}(d\theta)=-d\mathcal{L}_{X_1}\theta=d\theta=-\omega.
$$

Given a vector field $X$ generating a one-parameter group of non-strictly canonical transformations,
 we know that there exists 
a real number $a$ such that $\mathcal{L}_{X}\omega=a\,\omega$ and then the vector field
$X+a\,X_1$  is locally-Hamiltonian, because 
$$
\mathcal{L}_{X+a\,X_1}\omega=a\,\omega -a\, \omega =0.
$$
That means that there exists a closed 1-form $\alpha$ such that 
$$i(X)\omega+a\, i(X_1)\omega=\alpha.$$ 
Conversely, given a closed 1-form $\alpha$ the preceding relation defines a vector field $X$ such that $\mathcal{L}_{X}\omega=a\,\omega$ and then $X$ generates
a one-parameter (local) subgroup of non-strictly canonical transformations $\Phi_\epsilon$ with valence $e^{a\epsilon}$.

\subsection{Canonoid transformations }
\indent

As indicated above the set of canonical transformations is a catalogue of transformations preserving the Hamilton form of the dynamical equation. However, these are not the only ones that may be relevant for the study of the dynamics.
Here, we shall be interested in a type of transformation that preserves the 
Hamiltonian character of a particular  given  Hamiltonian system: they  are called canonoid  transformations with respect to this particular Hamiltonian system
 \cite{SaletanCromer, LeM84, NeOT87}. 
 They can be useful for the given specific problem, but not for other Hamiltonian systems.  Of course, all the canonical 
 transformations are canonoid but the converse is not true. 

In geometric terms, given a  Hamiltonian vector field $\Ga\in {\mathfrak X}_{{\rm H}}(M,\omega)$  in a symplectic manifold 
$(M,\omega)$, i.e. there exists a function $H\in C^\infty(M)$ such that $i(\Gamma)\omega=dH$, a transformation 
$\Phi\in \Diff(M)$ is said to be canonoid with respect to $\Gamma$, or with respect to its Hamiltonian $H$,  
if the transformed field $\Phi_*\Ga$ is also Hamiltonian, that is, $\Phi_*\Ga\in \mathfrak{X}_{\rm H}(M,\omega)$.  
 Since $\Phi$ is a diffeomorphism, we have that the vector field  $\Phi_*\Ga$ is Hamiltonian  with respect to $\omega$ if 
 and only if $\Ga$ is  Hamiltonian  with respect to the transformed 2-form $\Phi^*(\omega)$ \cite{CaR88}, i.e. there exists a function  
 $H'\in C^\infty(M)$ such that 
\begin{equation}
  i(\Ga)\Phi^*(\omega) = d H' \,. \label{defcanonoid}
\end{equation}
This means that if $\Phi$ is a canonoid transformation for $\Ga$ then $\Ga$ admits a new and  different  Hamiltonian structure. Therefore, this vector field $\Gamma$ will be  a bi-Hamiltonian system, that is, it is
Hamiltonian with respect to two different symplectic structures: the original symplectic form  $\omega$ and the new one $\Phi^*(\omega)$. 

Canonical transformations, either strictly canonical (that satisfy $\Phi^*(\omega)=\omega$) or non-strictly canonical transformations (that satisfy $\Phi^*(\omega)=r\,\omega$,  with $0\ne r\in \mathbb{R}$), are canonoid with respect to any Hamiltonian in a trivial way. The converse property is also true and if
a given transformation is canonoid with respect to any Hamiltonian function, it is canonical. Even it is enough when 
  the transformation is canonoid with respect to a more reduced family of Hamiltonians (see \cite{CuS72}, \cite{CaR90} 
  and \cite{CGILV,S98} and references therein).

These concepts can be generalised  to locally-Hamiltonian systems instead of Hamiltonian ones: 
If  $(M,\omega,\Gamma)$  is a locally-Hamiltonian dynamical system, a diffeomorphism  $\Phi:M\to M$ 
 is  a canonoid transformation with respect to $\Gamma$ when $\Gamma $ is 
locally-Hamiltonian with respect to $\Phi^*\omega$, i.e. if and only if
\begin{equation}
  \mathcal{L}_{\Ga}\,\Phi^*(\omega) = 0 \,, \label{definfcanonid}
\end{equation}
i.e.
\begin{equation}
  d [i(\Ga)\Phi^*(\omega)]=0.\label{definfcanonoid2}
\end{equation}

If we consider not just one transformation $\Phi$ but  a one-parameter group of 
canonoid transformations $\Phi_\ep$, then  this family of transformations  is canonoid with respect to 
a Hamiltonian vector field $\Gamma_H$  
if
and only if its infinitesimal generator $X$ is such that there exists a function  $K\in C^\infty(M)$
\begin{equation}
i(\Gamma_H) \mathcal{L}_{X}\omega =dK,\label{definfcanonoid}
\end{equation}
as one easily sees 
from  $i(\Ga_H)\Phi_\ep^*(\omega) = d _\ep$  when  taking the derivative 
with respect to $\epsilon$ at $\epsilon=0$. 
Analogously, if $\Gamma$ is locally-Hamiltonian with respect to $\omega$,  then $X$ induces a family
of canonoid transformations of $\Gamma$ if and only if 
\begin{equation}
 \mathcal{L}_{\Ga} \mathcal{L}_{X}\omega = 0 \,. \label{definfcanonoidloc}
\end{equation}
Later we shall discuss further characterizations and properties of these transformations, but before that, we are going to introduce a generalization of symmetry and constant of motion that it happens to be closely related to canonoid transformations.

\section{Master symmetries}
\indent
 
A function $T$ in a symplectic manifold  is said to be a generator of constants of
motion of degree $m$ if it is not preserved by the dynamics but it generates a constant of the motion by taking $m$ times its time derivative in an iterative way:
$$
 \frac{d}{dt}T \ne 0\,,\dots\,,\ \frac{d^{m}}{dt^{m}}T \ne 0\,,\ 
 \frac{d^{m+1}}{dt^{m+1}}T = 0 \,. 
$$
Of course, for $m=0$ we recover the usual definition of constant of motion.

In differential geometric terms, if $m>0$ and the  dynamics is given by a vector field $\Gamma$, these conditions are 
\begin{equation}\label{masterfun}
\mathcal{L}_\Gamma T\ne 0\,,\dots,\ \mathcal{L}_\Gamma^{m}T\ne 0\,,\ \mathcal{L}_\Gamma^{m+1}T= 0\,.
\end{equation}

We can introduce a time dependent observable associated to $T$ 
$$A=\sum_{n=0}^m(-1)^n \frac{A_n}{n!}t^n,\quad{\rm with}\ A_n=\mathcal{L}_\Ga^n T,$$ 
that is conserved along the motion, in the sense that
$$\frac{d}{dt}A=\left(\mathcal{L}_\Ga+\frac{\partial}{\partial t}\right) A=0.$$

Similarly,  as a symmetry of the dynamics $\Gamma$  is a vector field  $Z$  such 
that $[Z,\Ga]   = 0$,  a vector field $Z$ that satisfies the following two properties
\begin{equation}\label{masterfielduno}
 [Z,\Ga]   \ne 0  \,,\qquad
 [\,[Z \,,\Ga] \,,\Ga] = 0  \,,
\end{equation}
is called a `master symmetry' or a generator of symmetries of degree one for $\Ga$. If $m>1$ and $Z$ is such that 
\begin{equation}\label{masterfieldm}
[Z\,,\Ga] \ne 0  \,,\ldots,\
[\cdots[\,[Z \,,\underbrace{\Ga] \,,\Ga],\ldots,\Ga}_m]  \ne  0\,,
\ [\cdots[\,[Z \,,\underbrace{\Ga] \,,\Ga],\ldots,\Ga}_{m+1}]  =  0
\end{equation}
then it is called a `master symmetry' or a generator of symmetries of degree $m$ 
for $\Gamma$ \cite{Da93,Fe93,Ra99,Smir96,FinFo02,Cas02,DaSo05,Ra12}. 
Last condition in (\ref{masterfieldm}) can also be written as
 $\mathcal{L}_{\Gamma}^{m+1}(Z)=0$, in complete analogy to (\ref{masterfun}). 
As we will see below, for a Hamiltonian dynamical system this analogy 
goes further.

Let us now consider a locally Hamiltonian dynamical system $(M,\omega,\Gamma)$.
Observe  that the relation $\mathcal{L}_\Gamma (i(X)\omega)=i(\mathcal{L}_\Gamma X)\omega$ can be generalised to 
\begin{equation}
\mathcal{L}_\Gamma^k (i(X)\omega)=i(\mathcal{L}_\Gamma^k X)\omega,\quad \forall k\in \mathbb{N},\label{genleib}
\end{equation} which can easily be checked by induction on the number $k$, because it is valid for $k=1$ and if assumed true for a given index $k$, then
$$
\mathcal{L}_\Gamma^{k+1} (i(X)\omega)=\mathcal{L}_\Gamma(i(\mathcal{L}_\Gamma^k X)\omega)=i(\mathcal{L}_\Gamma^{k+1} X)\omega.
$$ 
Using this property (\ref{genleib}) for  the Hamiltonian vector field $X_T$  associated  to the function $T \in C^\infty(M)$ we obtain 
$$\mathcal{L}_\Gamma^k (i(X_T)\omega)=i(\mathcal{L}_\Gamma^k X_T)\omega,
$$
and therefore if $T$ is the generator of constants of motion of degree $m$, 
then its associated Hamiltonian vector field
$X_T$ is a master symmetry of degree $m$, because if $\mathcal{L}_\Gamma^{m+1}T=0$, then
$d(\mathcal{L}_\Gamma^{m+1}T)=0$, and therefore from
$$
i(\mathcal{L}_\Gamma^{m+1}X_T)\omega=\mathcal{L}_\Gamma^{m+1}\left(i(X_T)\omega\right)=\mathcal{L}_\Gamma^{m+1}(dT)=d(\mathcal{L}_\Gamma^{m+1}T)=0,
$$
we obtain that $\mathcal{L}_\Gamma^{m+1}X_T=0$.
 We call $T$ the generator of such  Hamiltonian master symmetry.
The converse is not true, in general,  because in order for the 
Hamiltonian vector field $X_T$ to be  
a master symmetry of degree $m$, the preceding relation shows that 
it is enough to demand that $X_{\mathcal{L}_{\Gamma_H}^{m+1}T}=0$, or equivalently
\begin{equation}\label{hammassym}
d\,\mathcal{L}_{\Ga_H}^{m+1}\, T=0.  
\end{equation}

Recall that in the particular case of  a Hamiltonian dynamical system $(M,\omega,H)$, 
$$
\mathcal{L}_{\Gamma_H}^{k} T=\{\cdots\{T \,,\underbrace{H\} \,,H\},
\ldots,H\}}_k.
$$
And the previous property can be rephrased by saying that a Hamiltonian vector field $X_T$
is a master symmetry of degree $M$ if and only if
$$d\{\cdots\{T \,,\underbrace{H\} \,,H\},
\ldots,H\}}_{m+1}=0.
$$

Next we illustrate this situation with a simple example. The Hamiltonian $H$ and the  vector field $\Ga_H$  of the one dimensional free particle are given by
$$
  H = \frac{1}{2}\,p^2 \,,{\quad} \Ga_H = p\,\pd{}{q} \,. 
$$
Then $X = \partial/\partial p$ satisfies
\begin{equation}
\Bigl [p\,\pd{}{q},\pd{}{p}\Bigr]   = \pd{}{q}  {\quad}{\rm and}{\quad} 
\Bigl [p\,\pd{}{q},\pd{}{q},\Bigr]   =  0\,,\label{master}
\end{equation}
i.e. $\mathcal{L}_{\Gamma_H}^2(X)=0$. Therefore $X$ is a master symmetry of degree $m=1$ for $\Gamma_H$.

In this example the vector field is Hamiltonian, $X=X_G$,
with $G(q,p)=-q$, and the relation (\ref{master})
can be rephrased in terms of Poisson brackets
$$\{G,H\}=-p\not=0\,,{\qquad}\{\{G,H\},H\}=0,$$
i.e. 
$$\mathcal{L}_{\Gamma_H} G=-p\ne 0, \qquad \mathcal{L}_{\Gamma_H}^2 G=0,
$$
and consequently $G$ is the generator of a master symmetry of degree $m=1$.

\section{Infinitesimal canonoid transformations and constants of the motion}
\indent

In this section we shall describe different ways of characterizing one-parameter  groups
of canonoid transformations for Hamiltonian and locally-Hamiltonian systems by 
using their infinitesimal generators. 

Let $X\in \mathfrak{X}(M)$ be a vector field in a symplectic manifold $(M,\omega)$ with a flow made of canonoid 
transformations with respect to a Hamiltonian $H$,  then, as indicated before,
$i(\Ga_H)\mathcal{L}_X\omega$ is an exact 1-form and therefore
\begin{equation}
i(\Ga_H)\mathcal{L}_X\omega=dK\label{defK}.
\end{equation}
Now a contraction with $\Ga_H$ of both sides of (\ref{defK}) shows that such a function $K$ is a constant of the motion.
 In particular, when the flow of $X$ is made up of  non-strictly canonical transformations,  there exists a 
 nonzero real number $a$ such that  $\mathcal{L}_X\omega=a\,\omega$ holds
and the function $K$ turns out  to be $K=a\, H$. 

An equivalent way of characterizing canonoid transformations 
for a (locally-) Hamiltonian vector field in a symplectic manifold $(M,\omega)$ is 
 the following one:
 
\begin{proposition}
a) The vector field $X\in \mathfrak{X}(M)$ is  the infinitesimal generator of a group of canonoid transformations for a locally-Hamiltonian 
vector field $\Gamma\in  \mathfrak{X}_{\rm LH}(M,\omega)$ if and only if $[X,\Gamma]$ is 
locally-Hamiltonian,  $[X,\Gamma]\in \mathfrak{X}_{\rm LH}(M,\omega)$.\hfill\break
b) Analogously $X$ is the infinitesimal generator of a group of canonoid transformations for a Hamiltonian $H$ if and only if $[X,\Gamma_H]$ is a Hamiltonian vector field, $[X,\Gamma_H]\in  \mathfrak{X}_{\rm H}(M,\omega)$. 
In this case its Hamiltonian function is
$\mathcal{L}_XH-K$, where $K$ is like in (\ref{defK}).
\end{proposition}

Proof.- \  a) We first compute $i([X,\Gamma])\omega$ and show that it is a closed form if and only if $X$ generates 
canonoid transformations.  In fact, if  $\Gamma\in  \mathfrak{X}_{\rm LH}(M,\omega)$, then $i(\Gamma)\omega$ is a closed form and
\begin{equation}
i([X,\Gamma])\omega= \mathcal{L}_X[i(\Gamma)\omega]-i(\Gamma)\mathcal{L}_X\omega=d\left(\omega(\Gamma,X)\right)-                                                                                      
i(\Gamma)\mathcal{L}_X\omega.\label{iXG}
\end{equation}
 Therefore,
$[X,\Gamma]$ is a locally-Hamiltonian vector field if and only if
$i(\Gamma)\mathcal{L}_X\omega$ is a closed 1-form,  or, equivalently, 
if and only if $X$ is an infinitesimal canonoid transformation for $\Gamma$.

b) Computing now $i([X,\Gamma_H])\omega$ as before we can conclude that $i([X,\Gamma_H])\omega$
is exact if and only if $i(\Gamma_H)\mathcal{L}_X\omega$ is exact. Moreover, in this case 
assuming  that
$i(\Gamma_H)\mathcal{L}_X\omega=d K$, as $\omega(\Gamma,X)=dH(X)=XH$, 
(\ref{iXG}) for $\Gamma=\Gamma_H$ reduces to 
$$i([X,\Gamma_H])\omega=d(\mathcal{L}_XH)-dK,$$
which proves that $[X,\Gamma_H]\in  \mathfrak{X}_{\rm H}(M,\omega)$, and 
its Hamiltonian function is  $\mathcal{L}_XH-K$.   {\hfill$\square$}
\\qed

As an immediate consequence of the previous proposition we see that any infinitesimal symmetry of the dynamics is the infinitesimal  generator 
of a family of canonoid transformations.

Cohomological techniques have been used  \cite{CI88} for studying non-canonical groups of transformations. 
In this paper, in  order to further study infinitesimal canonoid transformations
of a locally-Hamiltonian dynamical system $(M, \omega, \Gamma)$
and their relations with symmetries, 
we find useful to consider the differential operator of degree $-1$  on $\forms^\bullet(M)$
given by the contraction with the dynamical vector field, i.e.
$$i(\Gamma):\forms^{k+1}(M)\rightarrow\forms^k(M),$$
which obviously satisfies $i(\Gamma)\circ i(\Gamma)=0$. 
This, together with the de Rahm differential, allows us to define
new twisted boundary and coboundary operators, namely:
\begin{equation*}
\partial_\Gamma:=i(\Gamma)\circ d\circ i(\Gamma)\,,
\qquad
d_\Gamma:=d\circ i(\Gamma)\circ d\,.
\end{equation*}
Clearly both satisfy $\partial_\Gamma\circ\partial_\Gamma=0$, $d_\Gamma\circ d_\Gamma=0$ 
and the following relations 
$$i(\Gamma)\circ\partial_\Gamma=\partial_\Gamma\circ i(\Gamma)=0,\quad 
d_\Gamma\circ d=d\circ d_\Gamma=0, \quad \mathcal{L}_\Gamma \circ d_\Gamma= d_\Gamma  \circ   \mathcal{L}_\Gamma, \quad \mathcal{L}_\Gamma \circ 
\partial _\Gamma= \partial_\Gamma  \circ   \mathcal{L}_\Gamma.$$
Further properties of these operators are discussed in the Appendix A.

The closed and exact forms in $d_\Gamma$ cohomology have interesting dynamical 
properties.  An example of this is contained in the following proposition:

\begin{proposition}
Given a locally-Hamiltonian dynamical system $(M,\omega, \Gamma)$, 
the zero cohomology group of $d_\Gamma$, 
$H^0_\Gamma(M)=Z^0_\Gamma(M)=\{f\in C^\infty(M)\mid d_\Gamma f=0\}$ 
coincides with the set of generators of Hamiltonian dynamical symmetries.
\end{proposition}

Proof.- \  Let be $f\in C^\infty(M)$ and $X_f$ its Hamiltonian vector field, i.e. 
\begin{equation}\label{hvf}
  df=i(X_f)\omega.
\end{equation} 
Then, taking the Lie derivative in both sides of (\ref{hvf}),  we have  
\begin{equation}
 \mathcal{L}_\Gamma\,df=i([\Gamma,X_f])\omega + i(X_f)\,\mathcal{L}_\Gamma\omega \,,  \label{lgdf}
\end{equation} 
from where using that $\Gamma$ is a locally Hamiltonian vector field, and therefore it satisfies 
$\mathcal{L}_\Gamma\omega=0$, and the fact that $\omega$ is nondegenerate,
we arrive to $[\Gamma,X_f]=0$. That is,   $X_f$  is a dynamical symmetry  of $\Gamma$ if and only if $\mathcal{L}_\Gamma df= d_\Gamma f=0$,
i.e. $f\in Z_\Gamma^0(M)$.
\\qed

The generators of master symmetries can be characterized similarly. 
The result is that $G$ is the generator of a master symmetry of degree $m$ of $\Gamma$
if and only if $\mathcal{L}_\Gamma^m G\in Z^0_\Gamma(H)$, because 
$$d\mathcal{L}_\Gamma^{m+1}G=d\mathcal{L}_\Gamma(\mathcal{L}_\Gamma^{m}G)=(d\circ i(\Gamma)\circ d)(\mathcal{L}_\Gamma^{m}G)=d_\Gamma(\mathcal{L}_\Gamma^{m}G).
$$

The space of exact 1-forms with respect to $d_\Ga$, 
$B_\Ga^1(M)=\{d_\Ga f \mid  f\in C^\infty(M)\}$,
has also its dynamical interpretation.

\begin{proposition}
For a given locally-Hamiltonian dynamical system,
$(M,\omega,\Gamma)$, the vector field associated to  the $d_\Gamma$-exact 1-form
$\beta=d_\Gamma f$ is
$X_\beta=[\Gamma, X_f]$, where $X_f$ is the Hamiltonian
vector field of $f$.
\end{proposition}
Proof.- \ As  $\Gamma$ is a locally-Hamiltonian vector field, relation (\ref{lgdf}) reduces to 
$$i([\Gamma,X_f])\,\omega=\mathcal{L}_\Gamma d f=d_\Gamma f.$$
\\qed

Finally the space of closed 1-forms with respect to $d_\Ga$ is related  
to infinitesimal canonoid transformations as it is shown in the following
proposition:

\begin{proposition}
Let $(M,\omega,\Gamma)$ be a locally-Hamiltonian
dynamical system and  consider $\beta\in\forms^1(M)$. Then  $X_\beta\in\vfield(M)$ such that 
$i(X_\beta)\omega=\beta$ is the infinitesimal generator of a canonoid transformation
if, and only if, $\beta$ is $d_\Gamma$ closed, i.e.  $\beta\in Z^1_\Gamma(M)=\{\beta\in\forms^1(M)\mid  d_\Gamma\beta=0\}$.
\end{proposition}
Proof.- \  Recall  that $X_\beta$ induces a family
of canonoid transformations if, and only if, $\mathcal{L}_\Gamma(\mathcal{L}_{X_\beta}\omega)=0$,
or equivalently, $(d\circ i(\Gamma)\circ d\circ i(X_\beta) )\omega=0$, where $d\omega=0$
has been used. In terms of the twisted differential $d_\Gamma$ and $\beta$ 
the previous relation reduces to  $d_\Gamma \beta=0$.
\\qed

Note that this proposition is the translation to canonoid transformations and
twisted cohomology of the well known result about canonical transformations 
that are generated by closed forms in the de Rahm cohomology.
The 1-form $\beta$ is called the generator of the transformation. 

Next we shall consider infinitesimal canonoid transformations
which are master symmetries of degree $m$ of the dynamics.
We will show that for every such a transformation
we can  associate the generator of a Hamiltonian master symmetry of degree
$m-1$.

\begin{proposition}
Let $(M,\omega,\Gamma )$ be a locally-Hamiltonian dynamical system and assume
that $\beta\in \forms^1(M)$ generates an infinitesimal canonoid transformation for $\Gamma$. 
Then $X_\beta$ is a master symmetry of degree $m\ge 1$ for $\Gamma$ if and only if
$i(\Gamma) \beta$ is the generator of a Hamiltonian 
master symmetry of degree $m-1$.
\end{proposition}
Proof.- \  $X_\beta$ is a master symmetry of degree $m$ if and only if 
\begin{equation}\label{masterprop}
\mathcal{L}_\Ga^{m+1 }X_\beta=0,
\qquad
{\rm with}
\quad\ 
\mathcal{L}_\Ga^{m}X_\beta\ne0,
\end{equation}
and using that $\Gamma$ is Locally-Hamiltonian, $\mathcal{L}_{\Ga}\omega=0$, and the above mentioned property  (\ref{genleib}) for $k=m+1$ and $X=X_\beta$, i.e.
$$\mathcal{L}_\Ga^{m+1}\left( i(X_\beta)\omega\right)=i(\mathcal{L}_\Ga^{m+1}X_\beta)\omega,$$ 
together  with the definition of $X_\beta$, we see that 
(\ref{masterprop}) can be equivalently
written as 
$$\mathcal{L}_\Ga^{m+1 }\beta=0
\qquad 
{\rm and}
\quad
\mathcal{L}_\Ga^{m}\beta\ne 0.
$$
Now, if $X_\beta$ is an  infinitesimal canonoid transformation
$d_\Gamma\beta=0$, and  we have, 
$$0=\mathcal{L}_\Ga^{m+1}\beta=\mathcal{L}_\Ga^{m}( (i(\Gamma)\circ d+d\circ i(\Gamma))\beta) =(d\circ\mathcal{L}_\Ga^{m})( i(\Ga)\beta), 
\qquad {\rm for}\quad m\geq 1,$$ 
which, according to  eq. (\ref{hammassym}), is equivalent to say 
that $i(\Gamma)\beta$ is the generator of a Hamiltonian
master symmetry of degree $m-1$. 
\\qed

So far we have put into relation canonoid master symmetries with
Hamiltonian master symmetries of lower degree.
In the paragraphs below we shall go in the opposite direction,
namely we
shall relate canonoid transformations which are 
symmetries of the dynamics (recall that every symmetry of the dynamics is a canonoid
transformation) with generators of constants of motion of degree one.

With this aim we take, for a locally-Hamiltonian dynamical system 
$(M, \omega, \Gamma)$,
a 1-form $\beta\in Z^1_\Gamma(M)$, i.e. $d_\Gamma\beta=0$, and 
therefore,  $X_\beta$ is the infinitesimal generator
of a family of canonoid transformations. 
Assume that, at least locally, $\beta$ can be written
\begin{equation}\label{definitionofalpha}
\beta=\alpha+d G,
\end{equation}
where $\partial_\Gamma\alpha=0$ (it is easy to show that this
can always be achieved for the generators of symmetries
or around points where $\Gamma$ does not vanish,
see Appendix B).

Note that  $\partial_\Gamma\alpha =0$ implies  $i(\Gamma)(d(i(\Gamma)\alpha))=0$
and therefore the function 
$i(\Gamma)\alpha$ is a constant of motion. Moreover, (\ref{definitionofalpha}) shows that $d_\Gamma\alpha=0$. 

On the other hand,  $\mathcal{L}_\Gamma \alpha$ is a closed 1-form
because  $d\mathcal{L}_\Gamma \alpha=d_\Gamma \alpha=d_\Gamma \beta=0$.

The connection between canonoid symmetries and master symmetries
is expressed in the following proposition:

\begin{proposition}
If $X_\beta$ is the infinitesimal generator of a group
of dynamical symmetries of $\Gamma\in\mathfrak{X}_{\rm LH}(M,\omega)$, with
 $\beta=\alpha+d G$ and $\partial_\Gamma\alpha =0$, then 
$\mathcal{L}_\Gamma\alpha$ is exact, i.e. there exists a function $F$, uniquely defined up to 
addition of a constant,  such that 
\begin{equation}\label{definitionofF}
\mathcal{L}_\Gamma\alpha=dF,
\end{equation}
the function   $G$ is the generator of a constant
of motion of degree one and  the above function $F$  can be chosen such that  
\begin{equation}
\mathcal{L}_\Gamma G+F=0,\quad \mathcal{L}_\Gamma^2G=0.\label{GGF}
\end{equation}
\end{proposition}
Proof.- \ 
Consider the following equalities
\begin{equation}
i([X_\beta,\Gamma])\omega= -\mathcal{L}_{\Gamma} (i(X_\beta)\omega)             +
i(X_\beta)\mathcal{L}_{\Gamma}\omega=
-\mathcal{L}_{\Gamma} (\alpha+dG),
\end{equation}
we see that if $i([X_\beta,\Gamma])=0$ then $\mathcal{L}_{\Gamma} \alpha$ is exact and using the defining property for $F$ (\ref{definitionofF}), 
 the function $F+\mathcal{L}_\Gamma G$ is constant in every connected component of $M$.
With an adequate choice of the function $F$ in (\ref{definitionofF}), 
$F+\mathcal{L}_\Gamma G$ can be set to zero. Moreover, 
the function $F$ is a constant of motion, because
\begin{eqnarray*}
\mathcal{L}_\Gamma F=  i(\Gamma)\ dF=i(\Gamma)\mathcal{L}_\Gamma\alpha=
 \partial_\Gamma\alpha=0,
\end{eqnarray*}
and therefore $\mathcal{L}_\Gamma^2G=-\mathcal{L}_\Gamma F=0$, i.e. $G$ is the generator of a constant of the motion of degree one.
\\qed

This result can also be stated by saying that if the function $G$ and  the 1-form $\beta$ are related by 
(\ref{definitionofalpha}) and $X_\beta$ is a dynamical symmetry of $\Gamma$, then $X_G$ is a master symmetry of degree one.
If the dynamical system is Hamiltonian  with $\Gamma=\Gamma_H$ then  the relations (\ref{GGF}) are
\begin{equation}
\{G,H\}+F=0,\quad {\rm and} \quad  \{F,H\}=0.\label{GGF2}
\end{equation}

There is a kind of converse property. If $G$   is the generator of a constant of the motion of degree one, then the function  $F=-\mathcal{L}_\Gamma G$ is a constant of the motion. Now,
 for each 1-form $\alpha$ such that (\ref{definitionofF}) is satisfied we obtain that  $\partial_\Gamma\alpha=i(\Gamma)\mathcal{L}_\Gamma \alpha=0$ and
 $d_\Gamma\alpha=d(\mathcal{L}_\Gamma \alpha)=0$. Therefore,  the 1-form $\beta $  given by (\ref{definitionofalpha}) satisfies  $d_\Gamma\beta=0$, 
 and consequently, $X_\beta$ generates a one-parameter group of  canonoid transformations. 
Moreover, using the relations 
$$0=d(F+\mathcal{L}_\Gamma G)=\mathcal{L}_{\Gamma} (\alpha+dG)=
\mathcal{L}_{\Gamma} (i(X_\beta)\omega)    ,
$$
as $\Gamma$ is locally-Hamiltonian,  the preceding expression becomes
$$
0=\mathcal{L}_{\Gamma} (i(X_\beta)\omega   -
i(X_\beta)\mathcal{L}_{\Gamma}\omega=-i([X_\beta,\Gamma])\omega,
$$
hence, $[X_\beta,\Gamma]=0$, and $X_\beta$ is a canonoid  dynamical symmetry.

Note that the connection between (canonoid) dynamical symmetries
and generators of constants of motion of degree one is a generalisation of the strictly canonical case. In the latter $\alpha=0$, 
$F$ vanishes and the master symmetry of degree one is 
actually of degree zero, i.e. $G$ is a constant of motion.

To illustrate the previous results we can consider
the free particle in $\mathbb{R}$. The phase space $T^* \mathbb{R}$ is endowed with the canonical symplectic
 structure and the dynamics is $\Gamma=p\,\partial/\partial q$, with $H=p^2/2$. Let now $G$ be given by 
$$G=q\,p\,f(p), $$
and then in order for $F$ to satisfy the first relation  in (\ref{GGF2}),  $\{G,H\}+F=0$,
we must  choose  $F=-p^2\, f(p)$.  If the 1-form $\alpha$ is given by $\alpha=-(2f(p)+pf'(p))q\ dp$, 
that satisfy $i(\Gamma)\alpha=0$,  and therefore $\partial_\Gamma\alpha=0$,  then 
$\beta=\alpha+dG$
generates   a  canonoid
transformation 
$$X_\beta=f(p)\left(p\pd{}p+q\pd{}q\right),
$$
that clearly does not preserve (even up to a factor) the symplectic form.
Also notice that the previous expression (upon the addition of a Hamiltonian
vector field $g(p)\,\partial/\partial q$) is the most general
form for a dynamical symmetry 
of the one dimensional free particle. 

We should remark that the correspondence between (Hamiltonian) master 
symmetries and dynamical symmetries is not one to one. In fact, for the same 
master symmetry as before we can choose a different one form $\alpha$ 
and obtain a completely different symmetry. 
For instance, if we take $\alpha=-d(p\,q\, f(p))$ that satisfies all the required 
properties and the same function $G$ as before, we obtain $X=0$. 

Also the other way around: for any canonoid dynamical symmetry $X_\beta$,
as shown in the appendix, $\partial_\Gamma\beta=0$, which means that we 
can always take $G=0$ and the Hamiltonian master symmetry is trivial.

\section{Final comments}
In this paper we have studied, from a geometric perspective, 
the different transformations of a dynamical system that preserve 
 the Hamiltonian character of the  equations of motion.
We emphasize their similarities and discuss in depth
the case of canonoid transformations that are characterized
by preserving the structure of the equations for a 
particular Hamiltonian. 
This type of transformations include, in particular,
all the dynamical symmetries of the system. 

We present different intrinsic characterizations of the infinitesimal
generators for one-parameter groups of canonoid transformations and how 
they are related to canonical transformations.  A useful tool
for achieving this goal are certain {\it twisted}
homological and cohomological operators that are discussed in the paper. 

On the other hand we introduce a generalization of symmetries 
and constants of motion: the so called master symmetries and generators of 
constants of motion. The latter, actually, can be identified
with a conserved quantity that is polynomial in time. 
We establish two types of relations between master symmetries and
canonoid transformations that are symmetries of the dynamics.

An interesting point that could be worth studying is to try to extend the
relation between dynamical symmetries and canonoid transformations
to the case of master symmetries. Exactly as all dynamical symmetries are
canonoid transformations one could enlarge the class of transformations
so that they include all master symmetries of the system.

\section*{Appendix A}

Let $X\in\mathfrak{X}(M)$ be a vector field in a connected differentiable manifold. We define a map $d_X:\forms(M) \to\forms(M)$ as follows:
$$d_X:=d\circ i(X)\circ d= d\circ\mathcal{L}_X=\mathcal{L}_X\circ d.
$$

It is a degree one $\mathbb{R}$-linear map $d_X:\forms^\bullet(M)\to \forms^{\bullet+1}(M)$ such that:
\begin{itemize}
\item{} $d_X\circ d_X=0$ and $d\circ d_X=d_X\circ d=0$. 
\item{} $d_X\circ\mathcal{L}_X=\mathcal{L}_X\circ d_X$.
\item{} $d_X\circ i(X)+i(X)\circ d_X=\mathcal{L}_X^2$.
\item{} It is not a derivation but, for $\alpha, \beta\in \forms^r(M)$,  it satisfies 
$$d_X(\alpha\wedge \beta)= d_X\alpha\wedge \beta+ (-1)^r \alpha\wedge d_X\beta+d\alpha\wedge \mathcal{L}_X\beta+(-1)^r \mathcal{L}_X\alpha\wedge d\beta  \,.
$$
\end{itemize}

As $d_X\circ d_X=0$ we can define an associated cohomology where $B^0_X(M)$ is defined as $B^0_X(M)=\{0\}$ and $Z^r_X(M)$ and $B^r_X(M)$,  $r\in \mathbb{N}$, are given by 
$$ 
 Z^r_X(M)=\{\alpha\in \forms\,\!\!^r(M)\mid d_X\alpha=0\}, 
$$
and 
$$ 
 B^r_X(M)=\{\alpha\in \forms\,\!\!^r(M)\mid \exists \beta\in\forms\,\!\!^{r-1}(M) ,\ 
\alpha=d_X\beta\} \,, \quad r\geq 1.
$$

We remark that a consequence of the definition of $d_X$  
is the following chain of inclusions
$$B_X^r(M)\subset B^r(M)\subset Z^r(M)\subset Z^r_X(M).$$

We can characterize differently the space of $d_X$-closed and exact forms
$$
\begin{array}{rcl}
B_X^r(M)&=&\{\alpha\in \forms\,\!\!^r(M)\mid  \exists \beta\in B^{r}(M),\ \alpha ={\mathcal{L}}_X\beta \}  \\
Z_X^r(M)&=&\{\alpha\in \forms\,\!\!^r(M)\mid  \mathcal{L}_X\alpha\in Z^r(M)\}
\end{array}
$$
i.e. we can say that $B_X^r(M)$ is the image of $B^r(M)$ under $\mathcal{L}_X$
while $Z_X^r(M)$ is the preimage of $Z^r(M)$.

We have also introduced a degree -1 operator 
$\partial_X:\forms^r(M)\rightarrow\forms^{r-1}$
of the following form:
$$\partial_X:=i(X)\circ d\circ i(X)
=i(X)\circ\mathcal{L}_X
=\mathcal{L}_X\circ i(X).
$$
It satisfies
\begin{itemize}
\item{} $\partial_X\circ \partial_X=0$ and $i(X)\circ \partial_X=\partial_X\circ i(X)=0$. 
\item{} $\partial_X\circ\mathcal{L}_X=\mathcal{L}_X\circ \partial_X$.
\item{} $\partial_X\circ d=i(X)\circ d_X$ and $d\circ \partial_X=d_X\circ i(X)$.
\item{} $\partial_X\circ d+d\circ \partial_X=\mathcal{L}_X^2$.
\item{} $\partial_X\circ d_X+d_X\circ \partial_X=\mathcal{L}_X^3$.
\item{} For $\alpha,\beta\in \forms^r(M)$, we have 
$$\partial_X(\alpha\wedge \beta)= \partial_X\alpha\wedge \beta+ (-1)^r \alpha\wedge \partial_X\beta+i(X)\alpha \wedge \mathcal{L}_X\beta+(-1)^r \mathcal{L}_X\alpha\wedge i(X)\beta  \,.
$$
\end{itemize}

\section*{Appendix B}

Now we address the problem of existence of the {\it gauge fixing}  1-form
i.e. given $\alpha$ such that $d_\Ga\alpha=0$ does there exist
a function $f$ such that $\partial_\Gamma(\alpha+df)=i(\Gamma)di(\Gamma)(\alpha+df)=0$?
We have two partial positive answers to that question: a global one when 
$X_\alpha$ is a symmetry of the dynamics and
a local one around a point
in which $\Ga$ does not vanish. 

The first result is contained in the following proposition.

\begin{proposition}
Given a locally-Hamiltonian dynamical system
$(M,\omega,\Gamma)$, if the 1-form  $\alpha$ is such that $[X_\alpha,\Ga]=0$,
then $d_\Gamma\alpha=0$ and $\partial_\Gamma\alpha=0$.
\end{proposition}
Proof.- \ The relation
$$
\mathcal{L}_\Gamma\alpha=\mathcal{L}_\Gamma(i(X_\alpha)\omega)=i([\Gamma,X_\alpha])\omega
$$ shows
that if $X_\alpha$ is a symmetry we have $\mathcal{L}_\Ga\alpha=0$ and
applying to this identity the operator $d$ or $i(\Gamma )$
we obtain both results, because
$$
d(\mathcal{L}_\Gamma\alpha)=d_\Gamma\alpha=0,\quad i(\Gamma)\mathcal{L}_\Gamma\alpha=\partial_\Gamma\alpha=0.
$$
\\qed

The local result is made more precise in the following proposition.

\begin{proposition}
Let $p$ be a point in a Hamiltonian dynamical system 
$(M,\omega,H)$ such that $(\Gamma_H)_p\not=0$
and $\alpha$ any 1-form in $M$. Then, there exists a function  $f$, 
locally defined around $p$, such that $\partial_{\Gamma_H}(\alpha+df)=i(\Gamma_H)di(\Gamma_H)(\alpha+df)=0$.
\end{proposition}
Proof.- \  Note first that if the function 
 $f$  is such that 
\begin{equation}\label{gaugefixing}
 \mathcal{L}_{\Gamma_H} f= -i(\Gamma_H)\alpha
\end{equation}
applying $i_(\Gamma_H)\circ d$ to both sides we obtain $\partial_{\Gamma_H}(\alpha+df)=0$. 
 
But using the straightening out   theorem (see e.g. \cite{CrampPirani},\cite{AbrMarsdenRatiu}), if $\Gamma _H$ is different from zero at the point $p$, 
(\ref{gaugefixing}) can be transformed into an
explicit first-order ordinary differential equation around $p$,
whose solution always exists locally. 
\\qed

As the previous proposition shows the difficulties for 
finding a  locally defined 1-form in the family satisfying locally
condition $i(\Gamma_H)\,d\,i(\Gamma_H)\alpha=0$ arise when the
dynamical vector field vanishes at one point. In this case we can exhibit
en example in which the equation (\ref{gaugefixing}) cannot be solved.

Consider the Harmonic oscillator in one dimension with Hamiltonian given by the function in the phase space $T^*\mathbb{R}$, endowed with its 
canonical symplectic structure $\omega_0$, 
$$H=\frac12(p^2+q^2),$$
and therefore
$$\Gamma_H=p\pd{}q-q\pd{}p.$$
Note that $dH$, and therefore also $\Gamma_H$, vanish at the point $(0,0)$.

Take the canonical 1-form $\alpha=\theta_0=p\, dq$. One easily sees that $i(\Gamma_H)\alpha= p^2$, $i(\Gamma_H)d\alpha=-dH$, and it shows that $d i(\Gamma_H)d\alpha=0$
but the equation (\ref{gaugefixing}) in this case reads
$$q\pd fp-p\pd fq=p^2,$$
and the smooth solution should satisfy
\begin{eqnarray*}
\pd fq(q, p)&=&q\, g(q,p)-p,\cr
\pd fp(q,p)&=&p\, g(q,p).
\end{eqnarray*} 
for some smooth function $g$.
This pair of equations cannot be solved around $p=q=0$ because
from them we get
$$0=\pd {}q\left(\pd fp\right)-\pd{}p\left(\pd fq\right)(q,p)=
1-q\pd gp(q,p)+p\pd{}gq(q,p)$$
and the right hand side does not vanish at $p=q=0$.

In this situation, however, instead of the stronger condition (\ref{gaugefixing}) we can satisfy the weaker condition $i(\Gamma_H)\,d\,i(\Gamma_H)\alpha=0$, i.e.
$$\Gamma_H(\Gamma_H f)=-i(\Gamma_H)di(\Gamma_H)\alpha,$$
or in other words
$$\left(p\pd{}q-q\pd{}p\right)^2 f= 2\,q\,p$$
which can be solved by $f(q,p)=-\frac12q\,p$.
The new equivalent 1.form
$$\alpha'=\alpha+df=\frac12(p\,dq-q\,dp),$$
satisfies $\partial_\Gamma \alpha'=0$.

\section*{Acknowledgments} We acknowledge the support from research projects MTM--2009--11154,  MTM--2010-12116-E, FPA--2009-09638 (MEC, Madrid)  
and DGA-E24/1,  DGA-E24/2 (DGA, Zaragoza).


\end{document}